\documentclass{article}
\usepackage{spconf,amsmath,graphicx}
\usepackage{booktabs}
\usepackage{comment}
\usepackage{amssymb} 
\usepackage{color}
\usepackage{caption}
\usepackage{booktabs, makecell}
\usepackage{colortbl}
\usepackage{siunitx}
\usepackage[accsupp]{axessibility}  
\usepackage{multirow}
\usepackage{tabularx}
\usepackage{enumitem}
\usepackage{verbatim}
\usepackage{hyperref}
\usepackage{array}
\usepackage{booktabs}
\usepackage{subcaption}
\usepackage{tcolorbox}

\newcommand{\mL}{\mathbf{L}}
\newcommand{\mH}{\mathbf{H}}
\newcommand{\mS}{\mathbf{S}}
\newcommand{\mD}{\mathbf{D}} 
\newcommand{\mG}{\mathbf{G}} 
\newcommand{\mI}{\mathbf{I}} 
\newcommand{\mU}{\mathbf{U}}      
\newcommand{\mLambda}{\mathbf{\Lambda}}
\newcommand{\calT}{\mathcal{T}}

\newcommand{\mX}{\mathbf{X}} 
\newcommand{\mQ}{\mathbf{Q}} 
\newcommand{\mK}{\mathbf{K}} 
\newcommand{\mV}{\mathbf{V}} 
\newcommand{\mP}{\mathbf{P}} 
\newcommand{\calM}{\mathcal{M}}
\newcommand{\mT}{\mathbf{T}}

\newcommand{\calN}{\mathcal{N}}

\newtcolorbox{mybox}[1][]{colback=blue!3!white, colframe=pink!75!white, fonttitle=\bfseries, title=#1}

\title{\lowercase{i}HDR: Iterative HDR Imaging with Arbitrary Number of Exposures}
\name{Yu Yuan, Yiheng Chi, Xingguang Zhang, Stanley Chan%
     \thanks{The work is supported, in part, by the National Science Foundation under the grants 2133032, 2030570, and Angular Encoded Imaging.}}

\address{School of Electrical and Computer Engineering, Purdue University}

\begin{document}
%
\maketitle
\begin{abstract}
High dynamic range (HDR) imaging aims to obtain a high-quality HDR image by fusing information from multiple low dynamic range (LDR) images. Numerous learning-based HDR imaging methods have been proposed to achieve this for static and dynamic scenes. However, their architectures are mostly tailored for a fixed number (e.g., three) of inputs and, therefore, cannot apply directly to situations beyond the pre-defined limited scope. To address this issue, we propose a novel framework, iHDR, for iterative fusion, which comprises a ghost-free Dual-input HDR fusion network (DiHDR) and a physics-based domain mapping network (ToneNet). DiHDR leverages a pair of inputs to estimate an intermediate HDR image, while ToneNet maps it back to the nonlinear domain and serves as the reference input for the next pairwise fusion. This process is iteratively executed until all input frames are utilized. Qualitative and quantitative experiments demonstrate the effectiveness of the proposed method as compared to existing state-of-the-art HDR deghosting approaches given flexible numbers of input frames.
\end{abstract}


\vspace{-0.4em}
\section{Introduction}
\label{sec:intro}

The dynamic range of natural scenes can be very wide, spanning several orders of magnitude from dazzling sunlight to faint starlight. Main-stream imaging systems can only capture a very narrow dynamic range, reflected in the fact that we often get pictures with over-exposed or under-exposed regions. To address this limitation, many high dynamic range (HDR) imaging techniques have been developed.  
The conventional strategy is to use low dynamic range (LDR) exposure bracketing with different exposure values (EV) to compose the final single HDR image \cite{Debevec_l997_HDR, Mertens_2007_Mertens}. However, with the presence of foreground object movements and camera displacements, these methods often suffer from ghosting artifacts. Recently, many learning-based ghost-free HDR methods have been proposed to address this issue \cite{Kalantari_2017_DHDR, Wu_2018_DeepHDR, Yan_2019_AHDR, Yan_2020_CVIUHDR, Yan_2020_NonlocalHDR, Chen_2022_AttHDR, Liu_2022_ContextHDR, Song_2022_TransHDR, Tel_2023_HDR, Chi_2023_SNRHDR, Yan_2023_UnifiedHDR, Niu_2021_HDRGAN, Yan_2023_DiffHDR}.

While these HDR imaging methods have demonstrated promising performance, their network architectures are mostly tailored for a fixed number of (typically three) input LDR images. \textbf{We argue that this is rather an arbitrary choice because real-world LDR exposure bracketing can involve frames of varying numbers}. In some cases, only two frames are available for fusion, which could already be sufficient if, for example, the dynamic range of the scene is not very large; in other cases, more than three frames can be obtained to potentially boost the fusion quality. Most existing HDR methods cannot be directly applied when inputs consist of various numbers of images other than three while fusing only three of all available images, which results in sub-optimal solutions. \cite{Prabhakar_2019_Exposure, Catley-Chandar_2022_FlexHDR} can work with a flexible number of inputs; however, their model design requires massive pooling operations and optical flow alignment, which are inefficient and subject to potential information loss.

To cope with the stated problems, we introduce a HDR imaging framework iHDR in this paper, featuring an iterative fusion strategy that comprises a ghost-free Dual-input HDR fusion network (DiHDR) and a domain mapping network (ToneNet). At each iteration, DiHDR leverages side information from an LDR image pair to estimate an intermediate HDR image, and ToneNet transforms it back to the non-linear input domain. The domain-aligned image is fused again with the remaining LDR images iteratively until all frames are utilized. The main contributions of our work are summarized as follows. 
\vspace{-0.5em}
\begin{enumerate}
	\item{We propose a novel HDR imaging framework capable of handling a flexible number of inputs. The core ideas encompass iterative pairwise fusion and learned domain mapping.}
    \vspace{-0.6em}
	\item{We propose leveraging the structure tensor and difference mask of input images to enhance the learning capability of the proposed HDR fusion network. We built a semi-cross attention transformer to utilize the side features.}
\end{enumerate}
\vspace{-3mm}
\begin{figure}[h]
  \centering
  \includegraphics[width=0.99\linewidth]{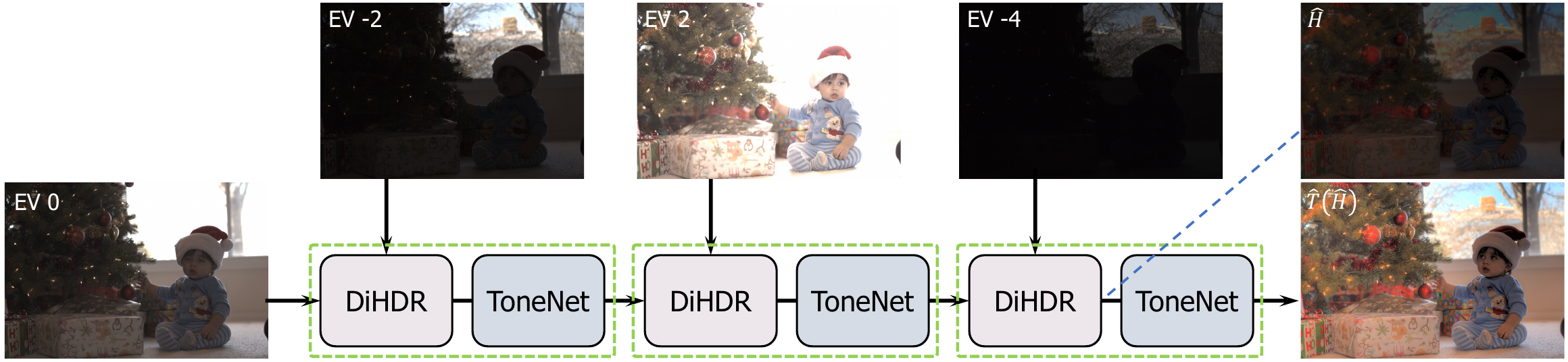}
  \caption{The proposed iHDR framework. ToneNet maps the linear HDR obtained from DiHDR back to the nonlinear domain consistent with the inputs.}
  \vspace{-0.8em}
  \label{fig:overview}
\end{figure}
\vspace{-1em}
\section{The Proposed Framework}
\label{proposed_framework}
\subsection{Problem Statement and Solution Overview}
\label{problem_statement}
Consider an HDR imaging problem where we want to convert a collection of LDR images $\left\{\mL_{1}, \mL_{2}, \ldots, \mL_{K}\right\}$ into a single HDR image $\mH$. Existing HDR methods often assume $K = 3$. As a result, if the method is learning-based, it will train a neural network with a fixed $K=3$. The question we want to ask here is the following

\begin{mybox}[Question]
Suppose we are given an exposure bracket with an \textbf{arbitrary $K (K \geq 2$) } number of LDR images, can we develop a generic learning-based framework that does not require re-training a new model?
\end{mybox}

To accomplish this goal, we propose a framework iHDR in Fig. \ref{fig:overview}. We highlight a few key features of this design:
\vspace{-0.5em}
\begin{enumerate}
\item \textbf{Iterative Pairwise Fusion.} The fusion is performed in a pairwise manner, hence allowing us to repeatedly blend LDR inputs for an arbitrary $K$. We design a dual-input HDR fusion network DiHDR to blend frames each time.
\vspace{-0.5em}
\item \textbf{Side Information $+$ Semi-cross Attention Transformer.} Based on image characteristics, we handcraft a collection of side information to assist the learning. This side information includes a pseudo-HDR image, structure tensor, and a difference mask. We designed a new module called the semi-cross attention transformer (SCAT) to integrate the side information.
\vspace{-0.5em}
\item \textbf{Learned Domain Mapping.} A tonemapping network ToneNet is introduced. ToneNet explicitly uses image sensor models instead of the standard $\mu$-law or pure learning-based methods to improve the sustained robustness of this mapping process.
\end{enumerate}

As illustrated in Fig. \ref{fig:overview}, for an LDR sequence with an arbitrary number of inputs, we first select a mid-exposed and sharp image as the reference image. Then, the remaining LDRs are sorted based on the difference in average luminance compared to the reference image, from smallest to largest. The reference image is initially fused with the non-reference image with the smallest average luminance difference through DiHDR. ToneNet aligns the domain of the intermediate HDR with LDR inputs, which then serve as the new reference image and are fused with the next non-reference image in the sorted order. This iterative process continues until all LDR images are effectively utilized, gradually increasing the details of the resulting HDR image. 

\vspace{-3mm}
\subsection{Side Information}
\begin{figure}[h]
  \centering
  \includegraphics[width=0.98\linewidth]{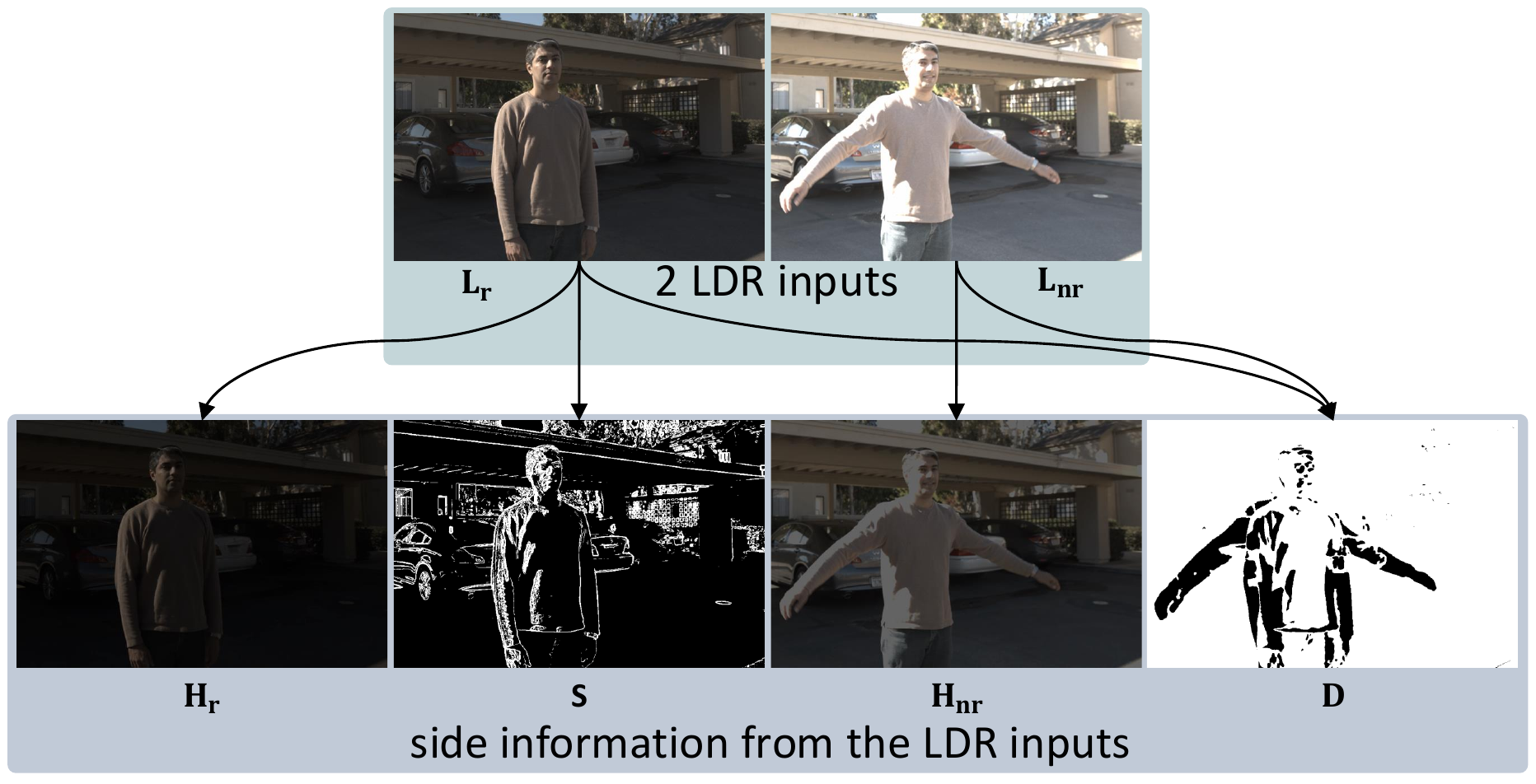}
  \caption{Three types of side information of inputs are utilized to enhance the learning capability of the network and suppress artifacts, including pseudo-HDR images of the inputs ($\mH_\text{r}$ and $\mH_\text{nr}$), the structure tensor of the reference frame ($\mS$), and the different mask between the two inputs ($\mD$).}
  \vspace{-0.8em}
  \label{fig: side information}
  
\end{figure}

To elaborate on our design, we describe a set of handcrafted features that will be used as the side information to guide the fusion. Fig. \ref{fig: side information} exhibits these features and their inputs.

\textbf{Pseudo-HDR Images.} Prior work such as Kalantari \textit{et al}. \cite{Kalantari_2017_DHDR} observed complementary capabilities of images in the LDR domain and the HDR domain. We construct the pseudo-HDR image from the $k$-th LDR image by
\vspace{-0.6em}
\begin{equation}
\mH_{k}=\frac{\mL_{k}^{\gamma}}{t_{k}}, \qquad k = 1,2,\ldots,K
\vspace{-0.6em}
\end{equation}
where $t_{k}$ is the exposure time for $\mL_{k}$, and $\gamma$ is typically 2.2.

\begin{figure}[h]
  \centering
  \includegraphics[width=0.99\linewidth]{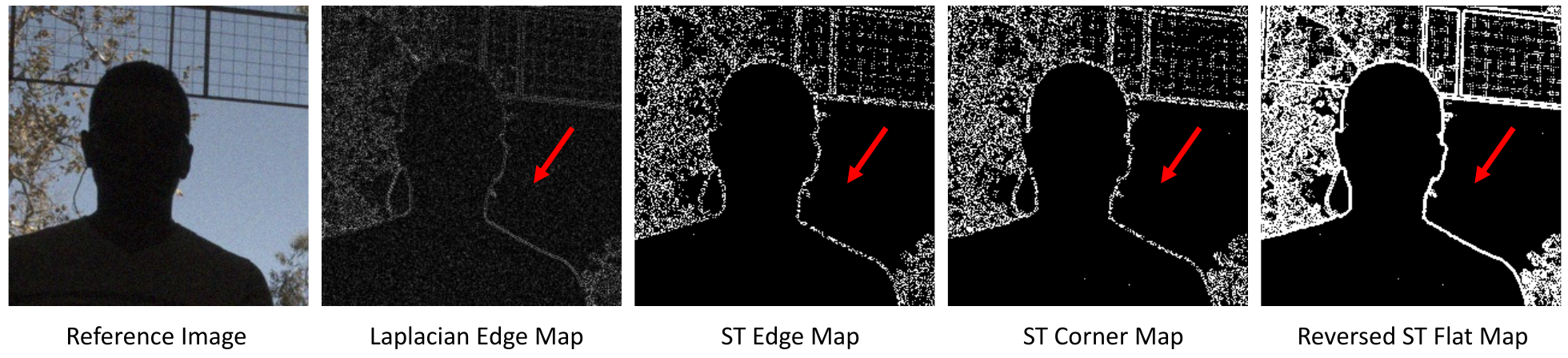}
  \caption{The reversed flat map of structure tensor has clearer textures and is more robust to noise.}
  \vspace{-0.8em}
  \label{fig:st}
\end{figure}

\textbf{Structure Tensor.} Some works \cite{Yan_2022_GradientHDR, Prabhakar_2016_MultiHDR} have noted that the gradient information of inputs can help eliminate artifacts. The structure tensor proposed in \cite{ Jung_2020_ST} has proven useful when extracting local gradients.  Consider a local neighborhood $\Omega$ (which is a square window containing $C$ pixels); we compute the horizontal and vector gradient to construct a $C \times$2 matrix $\mG$ followed by the eigendecomposition of the matrix $\mG^\intercal \mG$.
\vspace{-0.5em}
\begin{equation}
\mG=
\begin{bmatrix}
\nabla_x \mI, \quad \nabla_y\mI
\end{bmatrix}
\Rightarrow
\begin{bmatrix}
\mU, \quad \mLambda
\end{bmatrix}=\operatorname{eig}\left(\mG^\intercal \mG\right)
\vspace{-0.6em}
\end{equation}
The orientation of the gradient is captured by eigenvectors, and the magnitude of the gradient is captured by the eigenvalues. Based on the eigen-structure of $\mG^\intercal \mG$, we categorize the gradient information into three cases: edge, corner, and flat maps. In Fig. \ref{fig:st}, we show the conventional Laplacian edge map, which contains a substantially larger amount of noise. We utilize the reversed flat map $\mS$ to extract gradient information from the reference image, facilitating the network to mitigate artifacts. 

\textbf{Difference Mask.} Given a pair of LDR images, we denote one as the reference frame $\mL_{\text{r}}$ and the other as the non-reference frame $\mL_{\text{nr}}$. For the non-reference frame, we want to extract complementary features while minimizing any artifacts caused by motions and saturation. To this end, we introduce a difference mask to detect substantial changes occurring from $\mL_{\text{r}}$ to  $\mL_{\text{nr}}$. The difference mask $\mD$ has a formal definition below.
\vspace{-0.4em}
\begin{equation}
\mD =
\begin{cases}
1, & \text{if } \left|\calT(\mL_{\text{r}})-\calT(\mL_{\text{nr}})\right| > \text{threshold} \\
0, & \text{otherwise},
\end{cases}
\vspace{-0.4em}
\end{equation}
where $\calT(\cdot) = \texttt{rgb2gray}(\texttt{Blur}(\texttt{HistEq}(\cdot)))$ is the transformation. In our paper, we set the blur as a Gaussian blur with a radius of 7. The threshold is 0.2.

\vspace{-3mm}
\subsection{DiHDR and Semi-cross Attention Transformer}
\label{DiHDR_framework}
After introducing the side information, we discuss the ghost-free dual-input HDR fusion network DiHDR. The input to DiHDR is a collection of six elements \{$\mL_{\text{r}}$, $\mH_{\text{r}}$, $\mL_{\text{nr}}$, $\mH_{\text{nr}}$, $\mS$ and $\mD$\}. Fig. \ref{fig:iHDR} (a) illustrates the overall architecture of DiHDR. DiHDR is based on a U-shaped Transformer because of its remarkable performance in various image restoration tasks \cite{Zamir_2021_Restormer}. There are three main components of DiHDR:

\begin{figure*}[tb]
  \centering
  \includegraphics[width=1\linewidth]{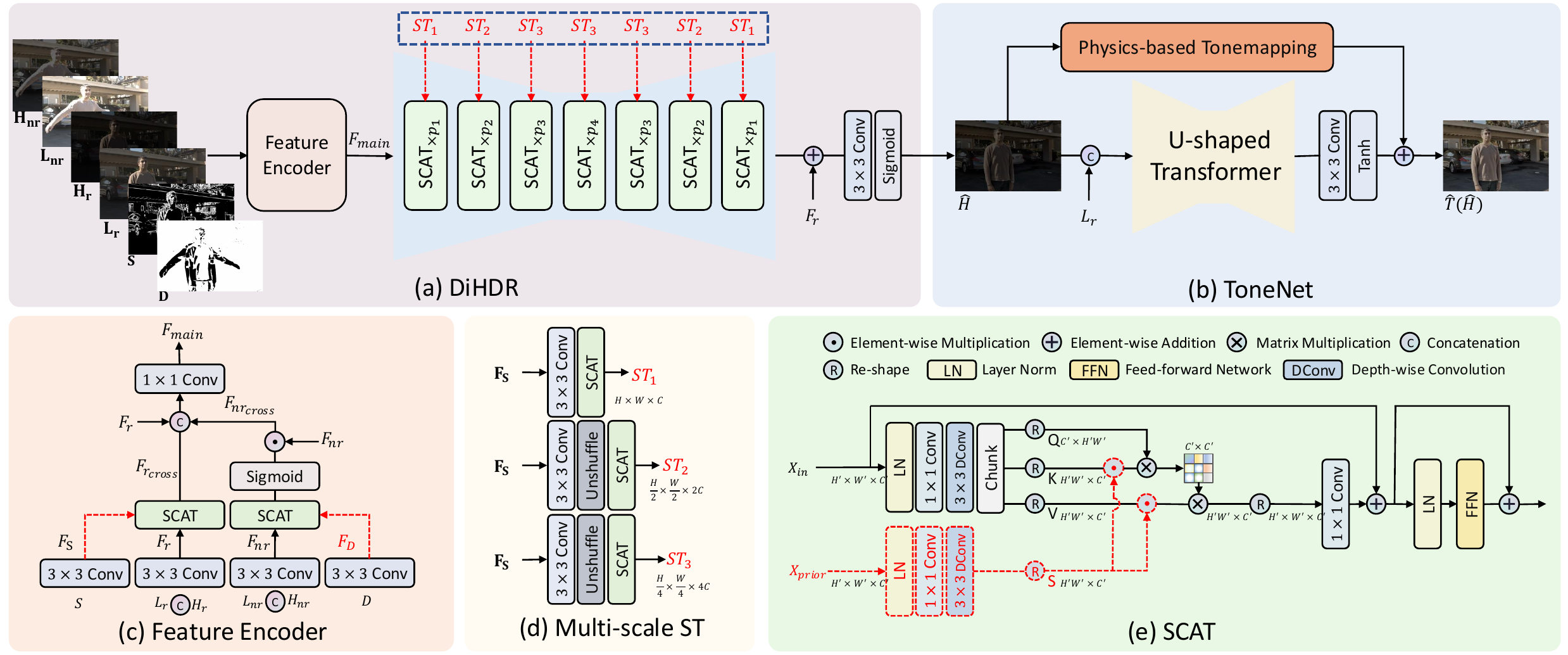}
  \caption{
  The overall framework of (a) DiHDR and (b) ToneNet. (c) Feature Encoder transforms the inputs and their side information into features using SCAT. (d) Multi-scale structure tensor priors are injected into the network to guide the structure of the generated image persistently. (e) In SCAT, additional prior features from structure tensor or difference mask are introduced into the transformer to capture cross attention. For SCAT blocks without prior inputs, the red dashed flows are masked.}
  \vspace{-0.8em}
\label{fig:iHDR}
\end{figure*}

\textbf{Feature Encoder.} 
The feature encoder that extracts and fuses multi-source features from the reference and non-reference branches.

\textbf{Multi-scale ST Guidance.} As illustrated in Fig. \ref{fig:iHDR} (d), we obtain three levels of structure tensor (ST) priors to guide the generation of HDR images towards the structural proximity of the target.

\textbf{Semi-cross Attention Transformer.} To effectively utilize prior information such as structure tensor and difference mask, we designed a semi-cross attention transformer (SCAT) shown in Fig. \ref{fig:iHDR} (e). SCAT is modified from multi-Dconv head transposed attention (MDTA) \cite{Zamir_2021_Restormer} overall. For the input feature $ \mX_\text{in} $, we apply layer normalization, 1$\times$1 convolution and 3$\times$3 depth-wise convolution before chunking into query ($\mQ $), key ($ \mK $), and value ($ \mV $). A crucial modification is the introduction of additional prior feature $ \mX_\text{prior} \in \mathbb{R}^{H^{\prime} \times W^{\prime} \times C^{\prime}} $. We project $\mX_\text{prior}$ onto $ \mP \in \mathbb{R}^{H^{\prime}W^{\prime} \times C^{\prime}} $ and perform L2 normalization along the spatial dimension. Subsequently, element-wise multiplications are performed between $\mP$ and $\mK$ and $\mV$. The  attention mechanism can be represented as
\vspace{-0.4em}
\begin{equation}
\text{Attention}(\mQ, \mK, \mV, \mP) = (\mP \odot \mV) \text{Softmax}\left(\frac{(\mP \odot \mK) \otimes \mQ}{\alpha}\right)
\vspace{-0.4em}
\end{equation}
where $\odot $ denotes element-wise multiplication, $\otimes $ represents matrix multiplication, and $ \alpha $ is a learnable scaling factor. 

\vspace{-3mm}
\subsection{ToneNet}
\label{tone_frame}
The goal of tonemapping in our proposed framework is to enable the iterative fusion scheme by using the fusion results from the previous step as input at the next step. This requires converting HDR images from the linear domain back to the nonlinear domain of the LDR input images. Images tonemapped by conventional methods such as $\mu$-law often vary with captured images significantly in brightness and contrast. Existing learning-based tonemapping methods also suffer from color biases and artifacts. We propose ToneNet, a physics-based mapping network shown in Fig. \ref{fig:iHDR} (b), to address this issue. We start from an imaging model and derive the following relation between a captured LDR image $\mL$ and the underlying HDR scene irradiance map $\mH$,
\vspace{-0.4em}
\begin{equation}
\mL = (c \cdot \mH) ^ {1/\gamma}
\label{eq: simplified imaging model}
\vspace{-0.5em}
\end{equation}
\noindent
where $c$ is an exposure-related scalar. More details can be found in the supplementary materials \url{https://sigport.org/documents/supplementary-materials-ihdr}.  

\vspace{-3mm}
\subsection{Loss Function}
In our framework, DiHDR and ToneNet can be trained together in an end-to-end manner, hence the training loss function is contributed by both. Following \cite{Kalantari_2017_DHDR}, $\mu$-law is used for calculating the loss of DiHDR
\vspace{-0.4em}
\begin{equation}
	\calM(\mH)=\frac{\log (1+\mu \mH)}{\log (1+\mu)}
\vspace{-0.4em}
\end{equation}
where $\mu$ is set to 5000. $\calM(\mH)$ is the $\mu$-law-based tonemapping result of $\mH$.

The loss function is defined as:
\vspace{-0.4em}
\begin{equation}
	\mathcal{L}=\|\calM(\widehat{\mH})-\calM(\mH)\|_{1}  +
 \lambda \|\widehat{\mT}- \mT\|_{1}
 \vspace{-0.4em}
\end{equation}
where $\mH$ denotes the ground truth HDR, $\widehat{\mH}$ represents the estimated HDR image by DiHDR. $\mT$ denotes the physics-based tonemapping results of ground truth HDR,  $\widehat{\mT}$ represents the estimated mapping results by ToneNet. $\lVert \cdot \rVert_1$
denotes $l_1$ loss. $\lambda$ indicates the weight of the mapping loss from ToneNet.

\vspace{-2.mm}
\section{Experiments}
\subsection{Implementation Details} 
\label{implementation_details}
\textbf{Datasets.} In this section, we perform extensive experiments to validate the effectiveness of the proposed approach. We use the dataset of SIG17 \cite{Kalantari_2017_DHDR}, where we split one set of three LDR images into two sets that share the same reference image. We generate 148 samples of training and 30 samples for testing dual-input frameworks. For the tonemapping labels required by ToneNet, we obtain them by manually adjusting the parameters of the physical model in Sec. \ref{tone_frame}. We also select 30 samples from SCTNet \cite{Tel_2023_HDR} as another test set, which has a wider variety of scenes and richer motion types. To evaluate the effectiveness of the proposed iterative fusion strategy, we use 10 labeled exposure brackets, each containing 9 dynamic scene frames with EVs of 0 and $\pm$\{1–4\}.
\begin{figure}[tb]
  \centering
  \includegraphics[width=1\linewidth]{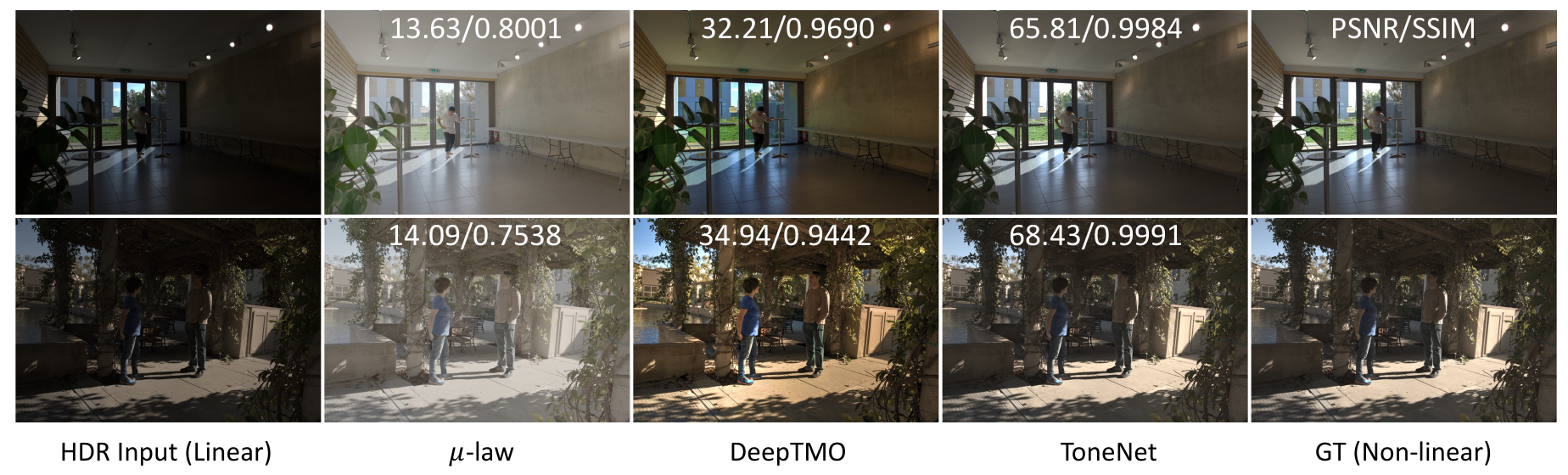}
  \caption{Tonemapping comparisons.}
  \vspace{-0.8em}
  \label{fig:tone}
\end{figure}
\textbf{Training Details.} The networks are trained with an AdamW optimizer with $\beta_{1}$=0.9 and $\beta_{2}$=0.999. The initial learning rate is set as 2$e$-4 and decreases to 1$e$-6 with the cosine annealing strategy. $ \lambda$ is set to 0.1. We train for 10,000 epochs with batch size 16 and set the input patch resolution as 128×128 pixels. The training process requires approximately 20 hours on 2 NVIDIA A100 GPUs.

\textbf{Metrics.} 5 metrics are used including: PSNR and SSIM evaluation in linear and tonemapped domains (by $\mu$-law), denoted as PSNR-$l$, PSNR-$\mu$, SSIM-$l$, SSIM-$\mu$, respectively, as well as HDR-VDP2 \cite{Mantiuk_2011_HDRVDP2}.

\vspace{-3mm}
\subsection{Comparison Experiments on HDR Tonemapping}
\label{comparison_tone_exp}
To verify ToneNet's capability of preserving the inherent HDR essence, we compare it with $\mu$-law and a learning-based tone-mapping, DeepTMO \cite{Rana_2020_DeepTone}. We fine-tune DeepTMO on SIG17 dataset to ensure fairness. Fig. \ref{fig:tone} illustrates the quantitative and qualitative results of these three methods. The results obtained through $\mu$-law suffer from saturation and contrast degradation, while DeepTMO tends to enhance colors excessively. ToneNet effectively preserves the original colors and contrast, which is crucial to ensure that biases in the tonemapping conversion do not accumulate significantly with the iterative fusion.
\vspace{-0.9em}
\subsection{Comparison Experiments on HDR Deghosting}
\label{comparison_deghosting_exp}
\begin{figure}[tb]
  \centering
  \includegraphics[width=1\linewidth]{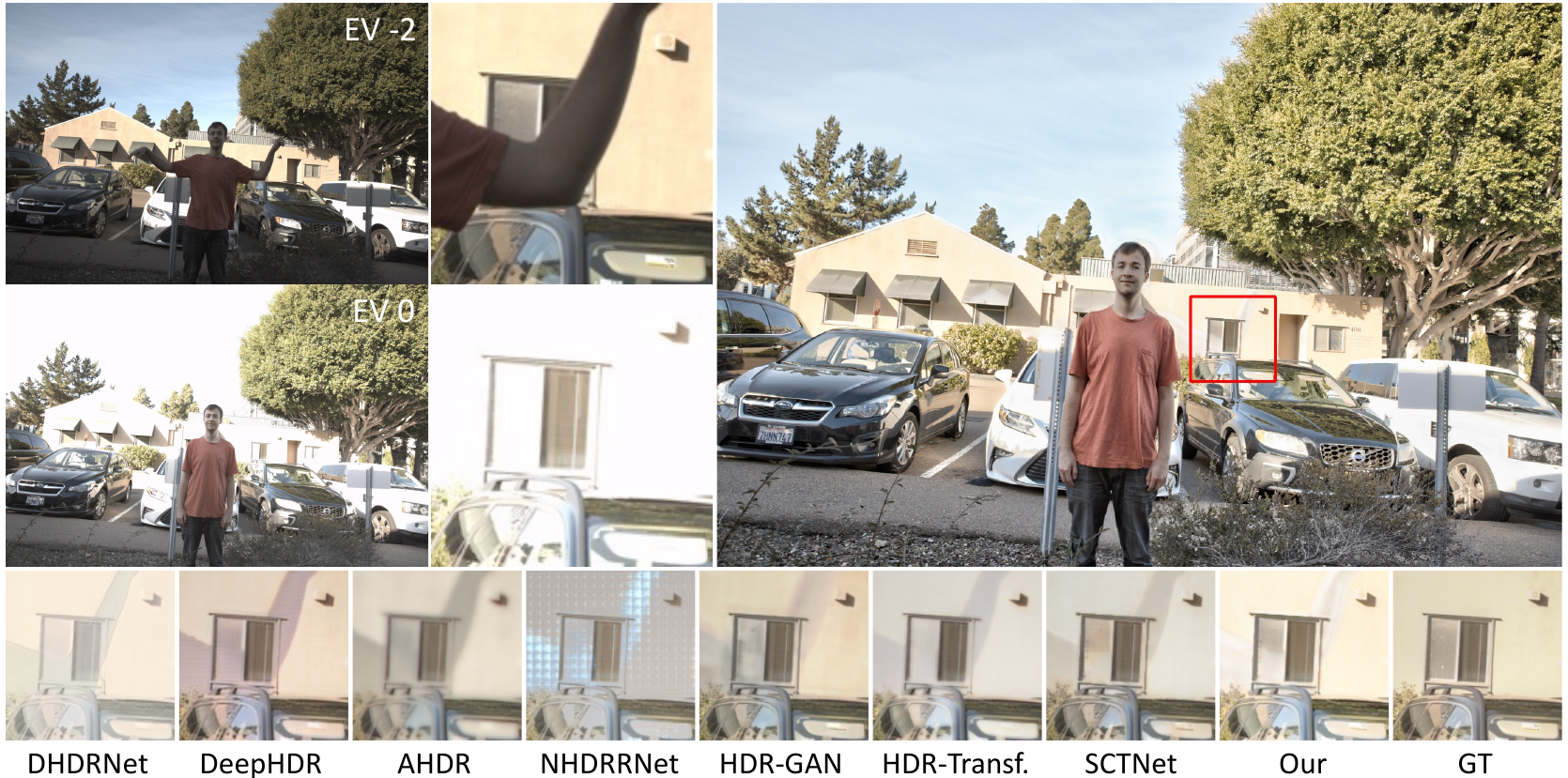}
  \caption{Qualitative comparison on SIG17 \cite{Kalantari_2017_DHDR} dataset. All methods are trained (fine-tuned) on the 2-input SIG17 dataset. \textit{Best viewed in zoom.}}
    \vspace{-0.8em}
  \label{fig:sig2}
\end{figure}

The comparison experiments are performed with 7 state-of-the-art HDR deghosting methods: DHDRNet \cite{Kalantari_2017_DHDR}, DeepHDR \cite{Wu_2018_DeepHDR}, AHDR \cite{Yan_2019_AHDR}, NHDRRNet \cite{Yan_2020_NonlocalHDR}, HDR-GAN \cite{Niu_2021_HDRGAN}, HDR-Transformer \cite{Liu_2022_ContextHDR}, SCTNet \cite{Tel_2023_HDR}.
To ensure fairness in comparing 2-input scenarios, we maintain the original structures and configurations of these networks and duplicate the non-reference frame to replace the original third LDR image as the input. These methods with 2-input are fine-tuned on the SIG17 dataset \cite{Kalantari_2017_DHDR}. As shown in Table. \ref{tab:sig17_2inputs}, our method outperforms the remaining seven methods across all five metrics. Fig. \ref{fig:sig2} also demonstrates that our approach has the weakest artifacts when handling two inputs. 
\begin{table*}[h]
\caption{\label{tab:sig17_2inputs}Quantitative comparison with different HDR deghosting methods on SIG17 dataset (2 LDR inputs). The default parameters for HDR-VDP2 are set to a one-meter viewing distance and a 21-inch display. Cells in red indicate the best, and blue denotes the second-best results.}
\centering
\scriptsize 

\begin{tabularx}{\textwidth}{l  *{8}{>{\centering\arraybackslash}X}}
\toprule
\multirow{2}{*}{Method}
 & DHDRNet & DeepHDR & AHDR  & NHDRRNet & HDR-GAN & HDR-Transf.  & SCTNet & Ours\\
&  \cite{Kalantari_2017_DHDR} & \cite{Wu_2018_DeepHDR}& \cite{Yan_2019_AHDR} & \cite{Yan_2020_NonlocalHDR} & \cite{Niu_2021_HDRGAN}  & \cite{Liu_2022_ContextHDR} & \cite{Tel_2023_HDR} & \\
\midrule
 PSNR-$l$ & 31.26 & 32.68 &  35.44 & 34.12 & 29.66 & \cellcolor{blue!10}36.64 & 34.16 & \cellcolor{red!10}38.75 \\
 PSNR-$\mu$ & 29.53 & 30.28 & 31.75 & 30.89 &  29.36 & 35.59  & \cellcolor{blue!10}36.48 & \cellcolor{red!10}41.63\\
  SSIM-$l$ & 0.8649 & 0.9407 & 0.9579 & 0.9573 & 0.8874 & 0.9543 & \cellcolor{blue!10}0.9669 & \cellcolor{red!10}0.9819 \\
 SSIM-$\mu$ & 0.9055 & 0.9203 & 0.9180 & 0.9673 & 0.9584 & \cellcolor{blue!10}0.9818 & 0.9774 & \cellcolor{red!10}0.9878 \\
  HDR-VDP2 & 56.95 & 57.73 & 59.45 & 51.19 & 59.87 & \cellcolor{blue!10}59.91 & 59.27 & \cellcolor{red!10}62.84 \\
\bottomrule
\end{tabularx}
\end{table*}

\begin{table*}[h]
\caption{\label{tab:our_dataset} Quantitative comparison with different HDR networks on our dataset.}
\centering
\scriptsize 
\begin{tabularx}{\textwidth}{l  *{8}{>{\centering\arraybackslash}X}}

\toprule
\multirow{2}{*}{Method}
 & AHDR  & NHDRRNet & HDR-GAN  & HDR-Transf.  & Ours & Ours & Ours & Ours  \\
& \cite{Yan_2019_AHDR} & \cite{Yan_2020_NonlocalHDR} & \cite{Niu_2021_HDRGAN} & \cite{Liu_2022_ContextHDR} & 2 frames & 3 frames & 5 frames & 9 frames \\
\midrule
  PSNR-$l$ & 13.06 & 11.03 & 15.28 & 16.27 & 13.92 & 14.71 & \cellcolor{blue!10}16.99 & \cellcolor{red!10}18.05 \\
  PSNR-$\mu$  & 13.84 & 14.13 & 17.58 &  15.54 & 15.84  & 16.71 & \cellcolor{blue!10}19.06 & \cellcolor{red!10}19.92\\
  SSIM-$l$  & 0.4261 & 0.4752 & 0.4969 & 0.6218 & 0.6777 & \cellcolor{blue!10}0.6806 &  \cellcolor{red!10}0.6823 & 0.6814 \\
  SSIM-$\mu$  & 0.4954 & 0.5174 & 0.6538 & 0.4606 & 0.5983 & 0.6577 & \cellcolor{blue!10}0.7765 & \cellcolor{red!10}0.7931 \\
  HDR-VDP2  & 47.00 & 48.44 & \cellcolor{blue!10}51.96 & 50.93 & 50.61 & 50.72 &  51.84 & \cellcolor{red!10}52.08 \\
\bottomrule
\end{tabularx}
\end{table*}
\vspace{-4mm}
\subsection{Flex Imaging.} 
\begin{figure}[tb]
  \centering
  \includegraphics[width=1\linewidth]{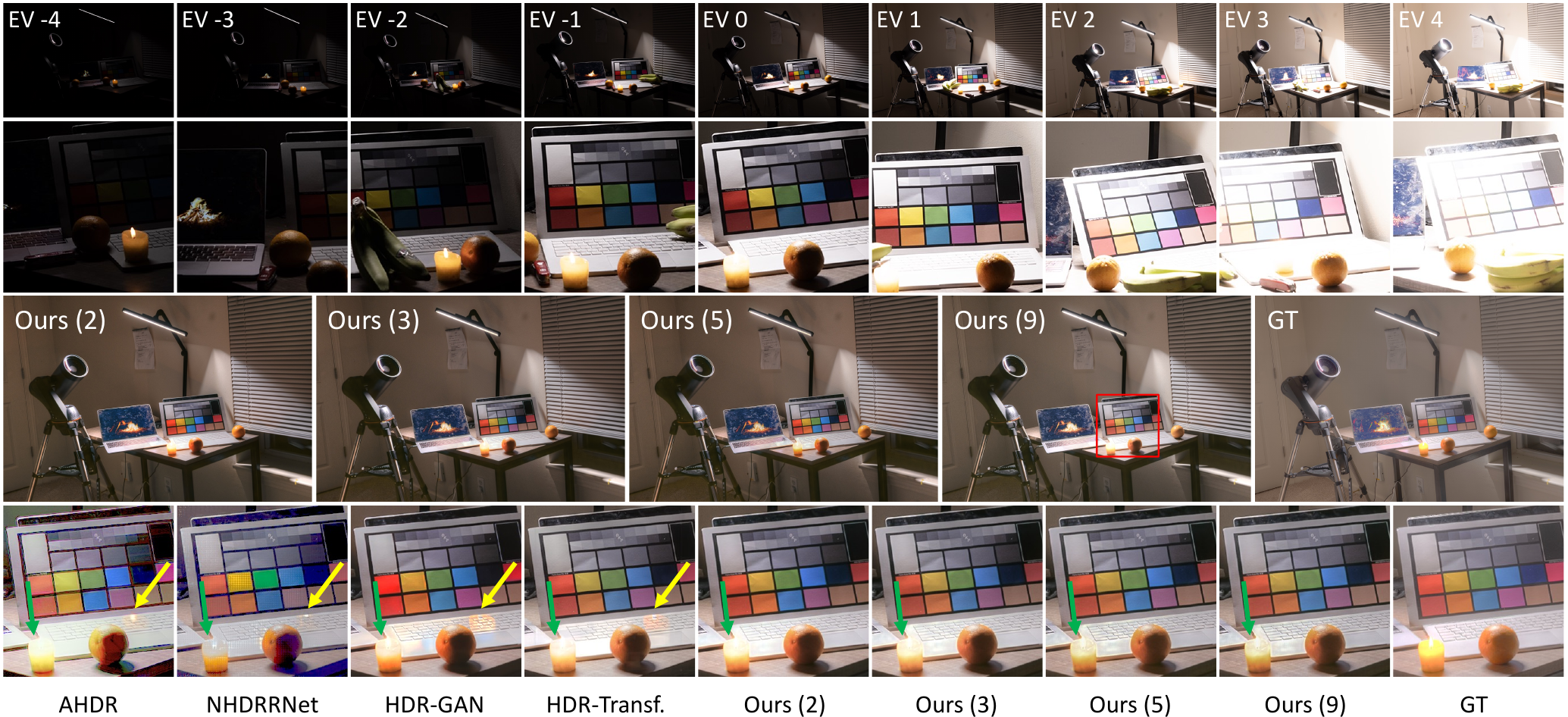}
  \caption{ Our method can accept a flexible number of LDR inputs, despite training with only two input images.}
  \vspace{-0.8em}
  \label{fig:9inputs}
\end{figure}
\begin{figure}
    \centering
    \small
    \vspace{-0.2em}
    \begin{tabular}{cc}
        \includegraphics[width=0.4\linewidth]{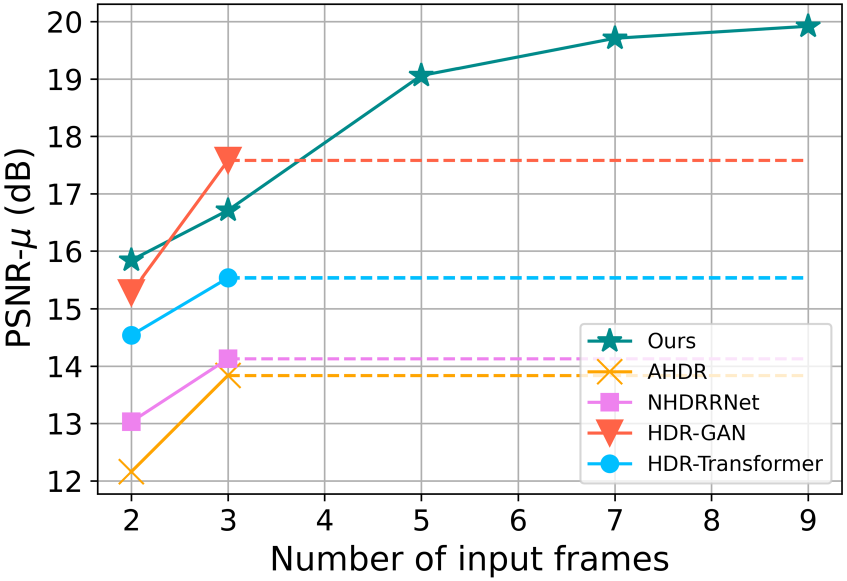} & \includegraphics[width=0.4\linewidth]{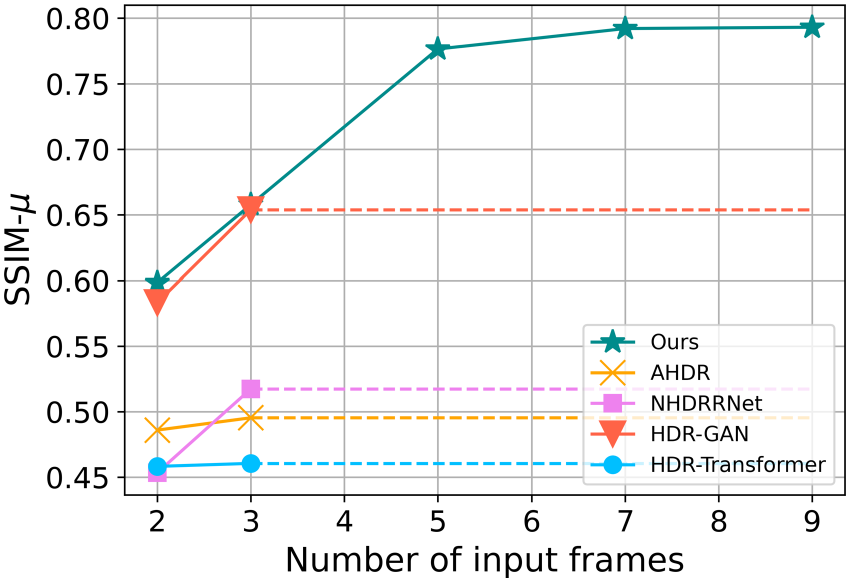} \\
        (a)  PSNR-$\mu$ curves & (b)  SSIM-$\mu$ curves
    \end{tabular}
    \vspace{-0.6em}
    \caption{Performance with varying input frame numbers.} 
    \label{fig:flx_number}
    \vspace{-0.8em}
\end{figure}

To validate the potential impact of the iterative fusion strategy, we conduct tests on our collected 9-input HDR dataset. As shown in Fig. \ref{fig:9inputs}, when the frames of input images increase, the result produced by our method exhibits higher dynamic range and more details. Fig. \ref{fig:flx_number} and Table. \ref{tab:our_dataset} illustrate that the PSNR-$\mu$ and SSIM-$\mu$ increase as the number of input images increases for our method. As the number of input frames increases, our method again presents superior performance to other three-input frameworks.
\vspace{-3mm}
\begin{table}[htbp]
  \centering
  \scriptsize
  \caption{\label{tab:ablation} Quantitative results of the ablation studies. All tests on a 2-input SIG17 dataset. $\mS$ denotes structure tensor and $\mD$ represents difference mask.}
 \vspace{-0.4em}
  \begin{subtable}[t]{0.48\linewidth}
    \centering
    \caption{$\mS$ and $\mD$}
    \begin{tabular}{ccc}
      \toprule
      Method & PSNR-$\mu$ & SSIM-$\mu$ \\
      \midrule
      W/o $\mS$ & 39.63 & 0.9753 \\
      Lap. Edge & 38.83 & 0.9534 \\
      W/o $\mD$ & 40.89 & 0.9796 \\
      Ours & \cellcolor{red!10}41.63  & \cellcolor{red!10}0.9878 \\
      \bottomrule
    \end{tabular}
  \end{subtable}
  \hfill
  \begin{subtable}[t]{0.48\linewidth}
    \centering
    \caption{SCAT Architecture}
    \begin{tabular}{ccc}
      \toprule
       Method  & PSNR-$\mu$ & SSIM-$\mu$ \\
      \midrule
      ${\text{RE}_{\{\mK, \mV\}}}$  & 40.60 & 0.9740 \\
      ${\text{SCAT}_{\{\mK\}}}$  & 41.11 & 0.9802 \\
      ${\text{SCAT}_{\{\mV\}}}$  & 41.36 & 0.9866 \\
      Ours  & \cellcolor{red!10}41.63 & \cellcolor{red!10}0.9878 \\
      \bottomrule
    \end{tabular}
  \end{subtable}
  \vspace{-0.3em}
\end{table}

\subsection{Ablation Studies}
\label{ablation_st}
\vspace{-0.2em}
\textbf{Structure Tensor and Difference Mask.} We conduct ablation experiments on the structure tensor ($\mS$) and difference mask ($\mD$). To show the effectiveness of the structure tensor, we replaced it with a Laplacian edge map and also compared the network with and without it. Regarding the difference mask, we remove the difference mask branch. The results in Table. \ref{tab:ablation} (a) indicate that directly injecting the Laplacian edge map may lead to some harm, while $\mS$ brings help. The presence of $\mD$ also contributes to performance improvement.

\textbf{Semi-cross Attention Transformer.} For the architecture design of SCAT, we explore different approaches: simply replacing the $\mK$ and $\mV$ branches with $\mP$ ($\text{RE}_{\{\mK, \mV\}}$);  only applying semi-cross attention to the $\mK$ branch ($\text{SCAT}_{\{\mK\}}$); only applying semi-cross attention to the $\mV$ branch ($\text{SCAT}_{\{\mV\}}$); and applying semi-cross attention to both the $\mK$ and $\mV$ branches (Ours). Table. \ref{tab:ablation} (b) results demonstrate that our proposed method is more effective at capturing dependencies between prior and input features.

\vspace{-3mm}
\section{Conclusion}
In this paper, we propose an iterative HDR imaging framework iHDR that can flexibly handle arbitrary number of input frames. In this framework, we design the DiHDR for pairwise fusion and the ToneNet for inter-domain transformation. We introduce SCAT, which effectively side information of input images to facilitate ghost-free fusion under dynamic scenes. Experimental results demonstrate that our method outperforms other methodologies when facing two and longer LDR bursts.

\clearpage
{
\bibliographystyle{IEEEtran} 
\bibliography{refs}
}

\newpage
\twocolumn[ 
    \begin{center}
        {\LARGE \textbf{Supplementary Material for iHDR:  Iterative HDR Imaging with Arbitrary Number of Inputs}} 
    \end{center}
    \vspace{5em} 
]

\section{Physics-based Tonemapping}
The goal of the physics-based tonemapping network is to enable the iterative fusion scheme by converting the HDR output of DiHDR from the linear domain back to the nonlinear domain of the LDR input images. Consequently, the tone-mapped results can serve as inputs for subsequent fusion steps.
Traditional tone-mapping methods, such as $\mu$-law, can map HDR images to nonlinear scales for display purposes. However, these tone-mapped images often exhibit significant variations from the captured images, particularly in terms of brightness and contrast. Moreover, learning-based tonemapping approaches are prone to color biases and visual artifacts. Such biases, even if minor, can accumulate and be exacerbated across fusion iterations, ultimately compromising the quality of the results.

 We address this issue by modeling the LDR imaging process. A realistic imaging model can be formulated as follows, similar to \cite{Chi_2023_SNRHDR, Qu_2024_2by2}. Consider an LDR image, $\mL$, captured at an exposure time of $t$ where the underlying HDR scene irradiance map is represented by $\mH$. 
\begin{align}
\setcounter{equation}{8}
\mL = \texttt{ADC} \Big\{ \xi  \times \texttt{Clip} \Big\{ 
& \text{Poisson} \big( t \times \text{QE} \times (\mH + \mu_{\text{dark}}) \big) 
\Big\} \nonumber \\ 
& + \calN(0, \sigma_{\text{read}}^2) \Big\}^{1/\gamma}
\label{eq:imaging_model}
\end{align}

 where $\xi$ is the conversion gain, QE is the quantum efficiency, $\mu_{\text{dark}}$ is the dark current, and $\sigma_{\text{read}}$ is the read noise standard deviation. Here, $\text{Poisson}$ represents the Poisson distribution characterizing the photon arriving process and the dark current effect, and $\calN$ represents the Gaussian distribution characterizing the sensor noise. $\texttt{ADC}\left\{\cdot\right\}$ is the analog-to-digital conversion and $\texttt{Clip}\left\{\cdot\right\}$ is the full well capacity induced saturation effect. We assume a linear camera response function for CMOS sensors and that the imperfections in the pixel array, ADC, and color filter array have been mitigated.
 
 Since our goal is to convert the estimated $\widehat{\mH}$ to the domain of $\mL$ while preserving essential HDR information, we can remove the random perturbations and lossy processes in the imaging model. For HDR datasets, we simplify the  parameters $\xi$, $t$, and QE by absorbing them into one exposure-related scalar $c=4.5$ (Equation. (5)), thus providing a physically motivated initial estimate.

\section{Datasets}
\textbf{Our 9-input LDR Dataset.} This paper collects 10 exposure brackets with labels, each containing 9 frames of dynamic scenes with EV values of $\pm$4, $\pm$3, $\pm$2, $\pm$1, and 0. We capture the data using a Sony $\alpha$6400 camera mounted on a tripod. The resolution of all images is downsampled to 1500 $\times$ 1000 pixels, as shown in Fig. \ref{fig:our_dataset}.

\setcounter{figure}{8}
\begin{figure}[tb]
  \centering
  \includegraphics[width=1\linewidth]{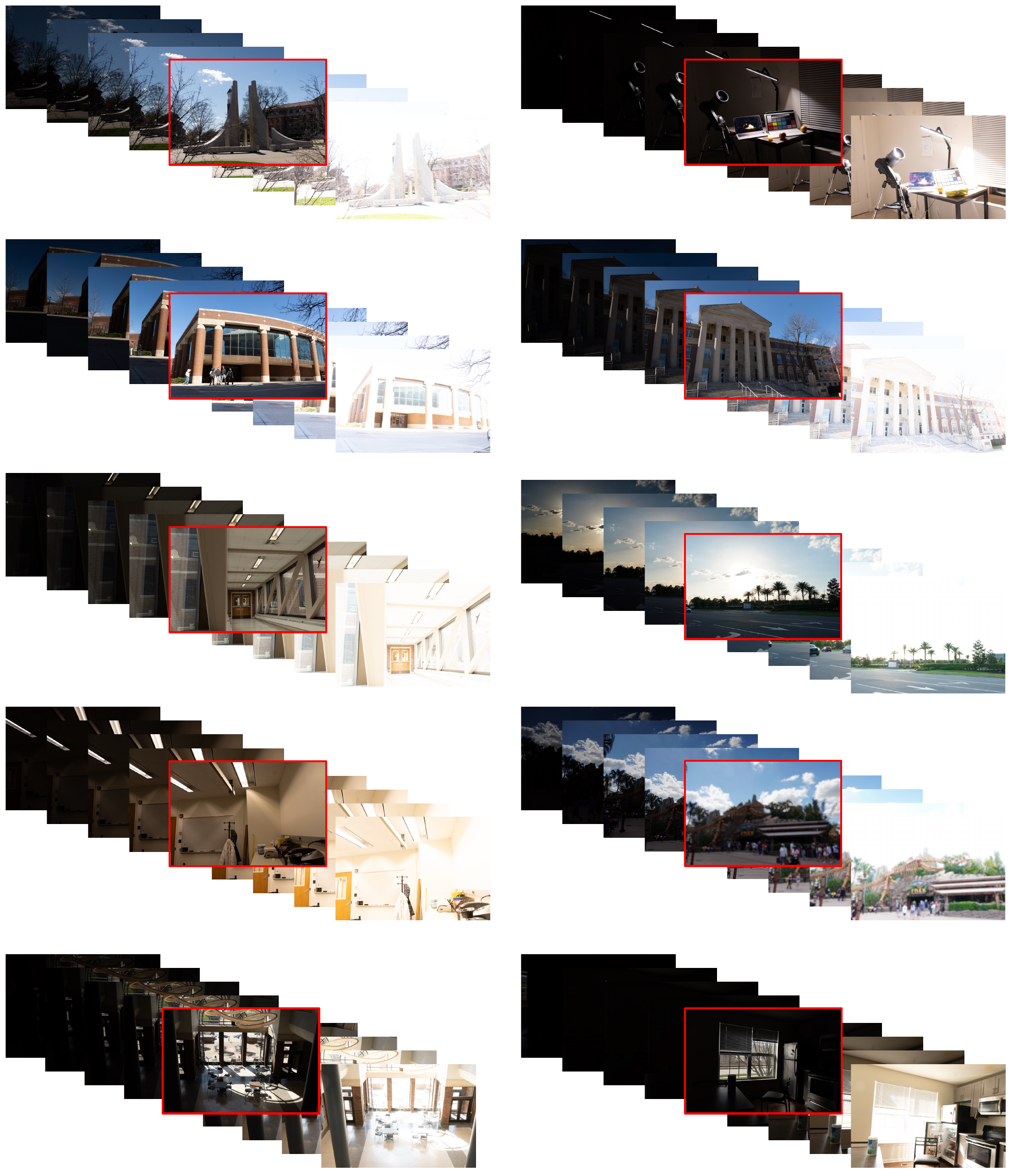}
  \caption{Test set generated by this paper. Our dataset highlights full motions and 9-frames from EV -4 to EV 4. The reference frame (EV 0) is marked by a red box in each subset.}
  \label{fig:our_dataset}
\end{figure}

\section{Computational Cost}
Table. \ref{tab:cost} presents a comparison of computational cost for various methods. The penultimate column lists the parameters of DiHDR, along with the wall time for processing two inputs. The last column provides the total time for DiHDR and ToneNet to process three input frames (2 DiHDR passes and 1 ToneNet pass)
Additionally, we compare GMACs for different methods. We observe that the computational time and complexity of iHDR is lower than or comparable to HDR-Transformer and SCTNet on 3 inputs of $1500\times1000$ images.

\setcounter{table}{3}
\begin{table}[t] 
  \caption{\label{tab:cost}Computational cost comparison of the proposed solution against other SOTA methods. The input size is set to 1000 × 1500 pixels. The speed is measured on an NVIDIA A100 GPU.}
  \scriptsize 
  \centering
  \begin{tabularx}{\columnwidth}{l *{6}{>{\centering\arraybackslash}X}} 
    \toprule
    Method & DeepHDR & AHDR & HDR-GAN & HDR-Transf. & SCTNet & iHDR \\
     & \cite{Wu_2018_DeepHDR}& \cite{Yan_2019_AHDR} & \cite{Niu_2021_HDRGAN}  & \cite{Liu_2022_ContextHDR} & \cite{Tel_2023_HDR} & \\
    \midrule
    GMACs & 1453.70 & 2166.69 & 778.81 & 981.81 & 293.77 &  374.76 \\
    Time (s)  & 0.29 & 0.35  & 4.85 & 6.86 & 7.12 &  6.93 \\ 
    \bottomrule
  \end{tabularx}
\end{table}

\section{Additional Visual Results}
\subsection{Experiments on HDR Deghosting.}

\begin{figure}[tb]
  \centering
  \includegraphics[width=1\linewidth]{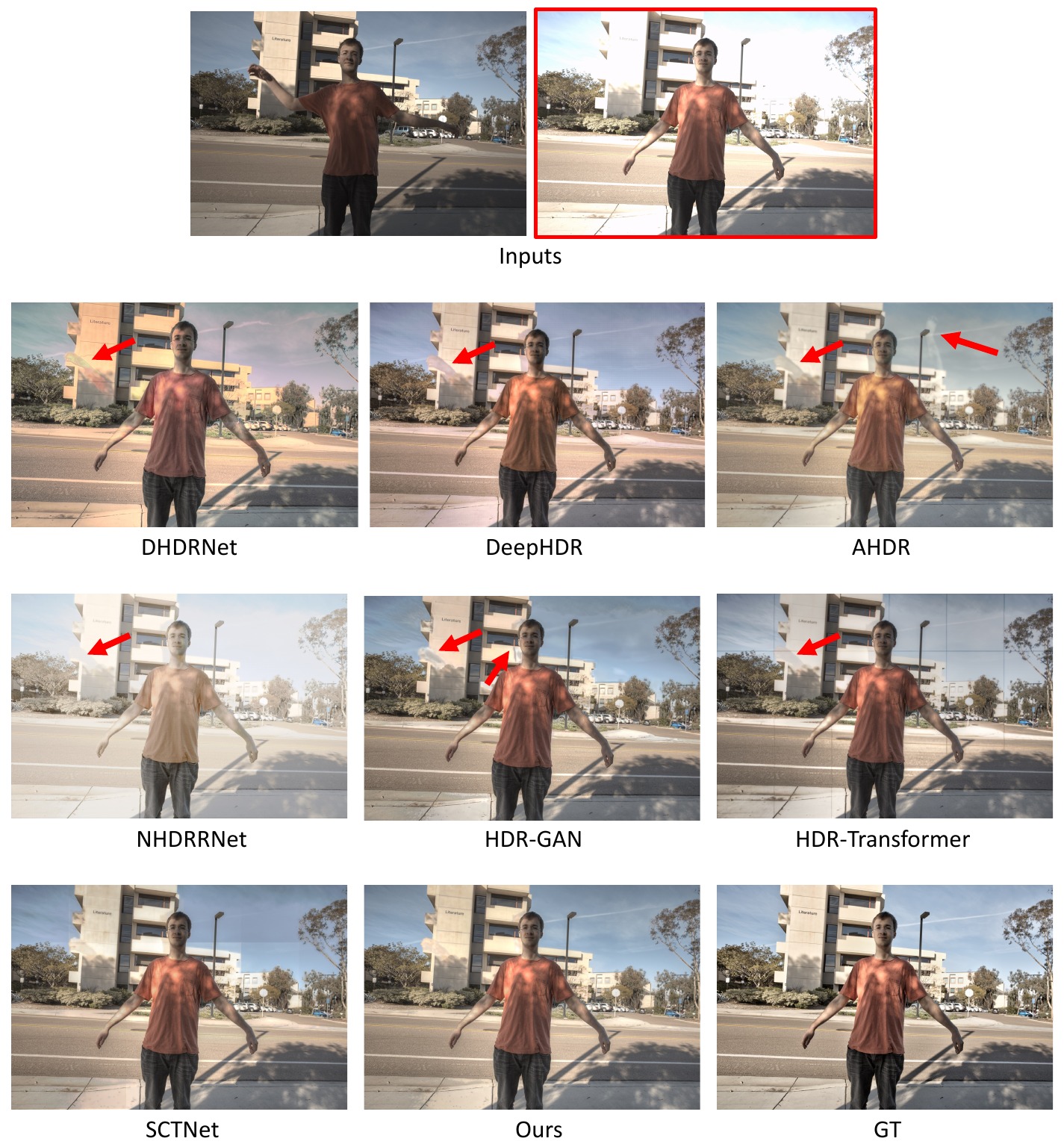}
  \caption{Qualitative comparison on \textit{Scene. \uppercase\expandafter{\romannumeral1}} from the SIG17 \cite{Kalantari_2017_DHDR} dataset. Results obtained by DHDRNet \cite{Kalantari_2017_DHDR}, DeepHDR \cite{Wu_2018_DeepHDR}, AHDR \cite{Yan_2019_AHDR}, NHDRRNet \cite{Yan_2020_NonlocalHDR}, HDR-GAN \cite{Niu_2021_HDRGAN}, HDR-Transformer \cite{Liu_2022_ContextHDR}, SCTNet \cite{Tel_2023_HDR} and Ours. The red box represents the reference frame.}
  \label{fig:sig_2}
\end{figure}

\textbf{Results on 2-input SIG17 Dataset.} Fig. \ref{fig:sig_2} show the results of HDR deghosting experiments.
Our method outperforms other methods in suppressing ghosting artifacts.

\begin{figure}[tb]
  \centering
  \includegraphics[width=1\linewidth]{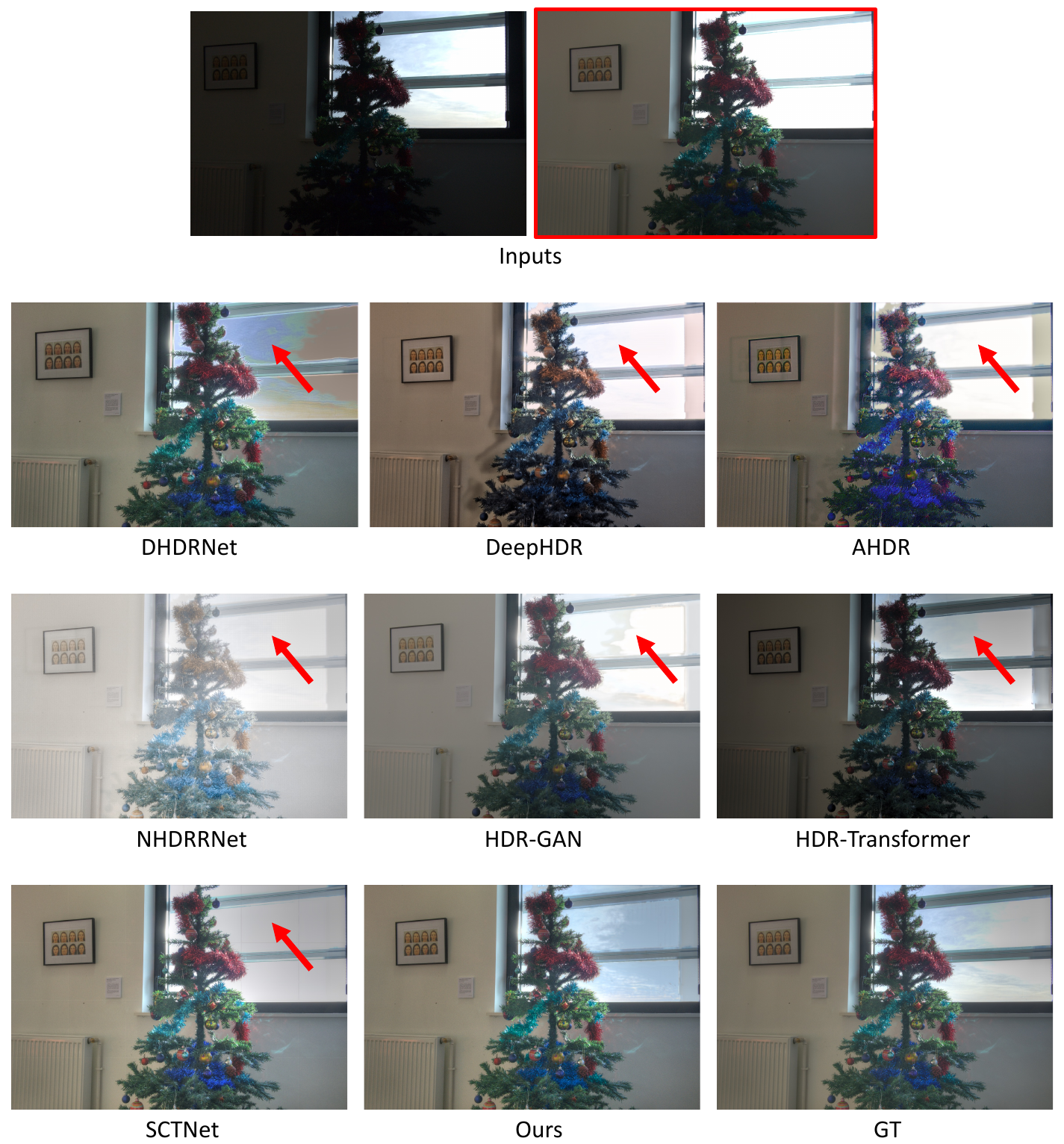}
  \caption{Qualitative comparison on \textit{Scene. \uppercase\expandafter{\romannumeral2}} from the SCTNet dataset\cite{Tel_2023_HDR}. Results obtained by DHDRNet \cite{Kalantari_2017_DHDR}, DeepHDR \cite{Wu_2018_DeepHDR}, AHDR \cite{Yan_2019_AHDR}, NHDRRNet \cite{Yan_2020_NonlocalHDR}, HDR-GAN \cite{Niu_2021_HDRGAN}, HDR-Transformer \cite{Liu_2022_ContextHDR}, SCTNet \cite{Tel_2023_HDR} and Ours. The red box represents the reference frame.}
  \label{fig:sct_3}
\end{figure}

\textbf{Results of Generalization Performance.}
Since all (2-input) HDR deghosting methods are trained on the 2-input SIG17\cite{Kalantari_2017_DHDR} dataset, we explore their generalization performance on the untrained SCTNet dataset\cite{Tel_2023_HDR}. Fig. \ref{fig:sct_3} demonstrate the results of these methods tested on out-of-domain data. Our method outperforms others in ghosting suppression, color reproduction, and detail preservation.

\subsection{Flex Imaging}
\begin{figure}[tb]
  \centering
  \includegraphics[width=1\linewidth]{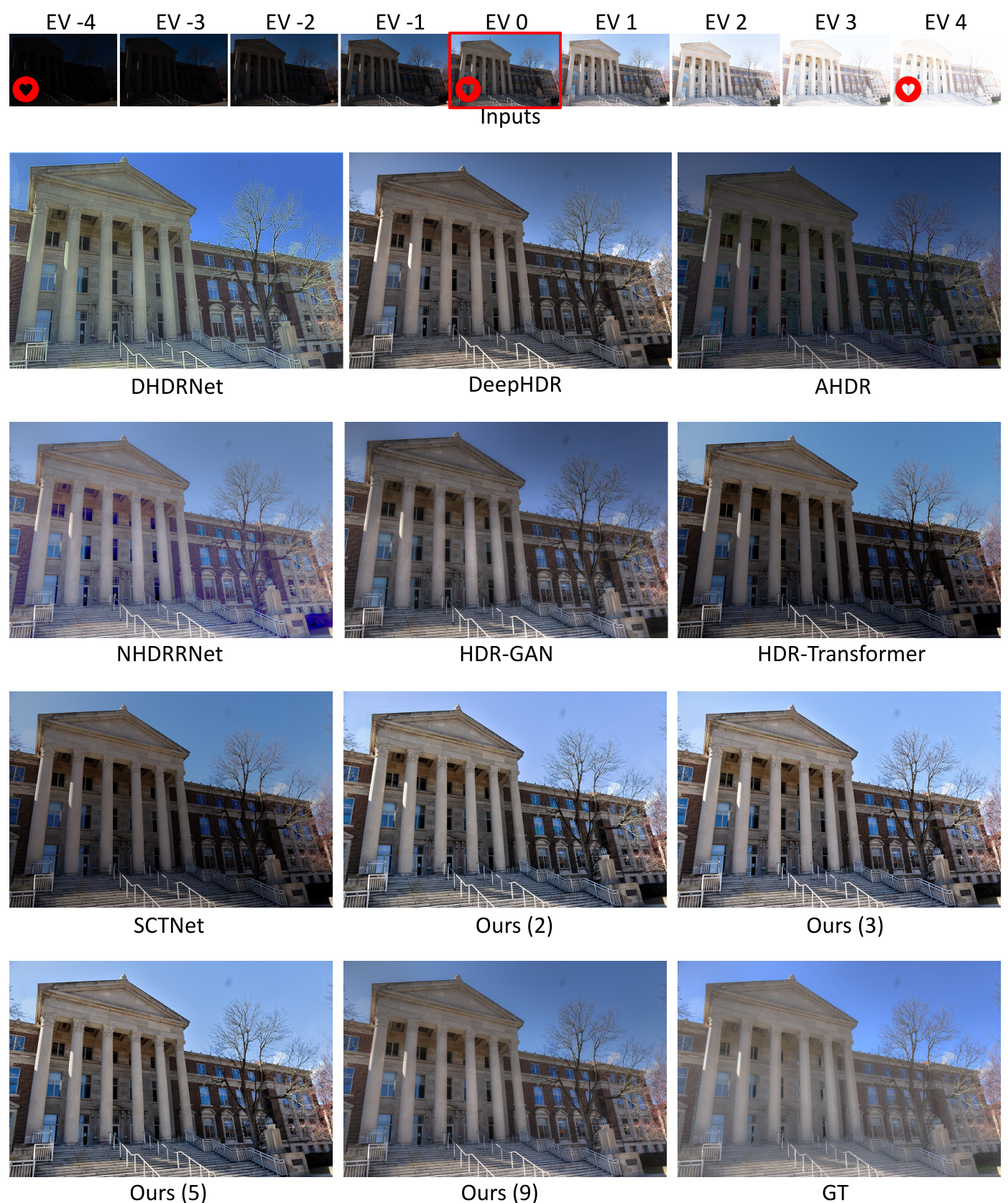}
  \caption{Qualitative comparison on \textit{Scene. \uppercase\expandafter{\romannumeral3}} from our dataset. Results obtained by DHDRNet \cite{Kalantari_2017_DHDR}, DeepHDR \cite{Wu_2018_DeepHDR}, AHDR \cite{Yan_2019_AHDR}, NHDRRNet \cite{Yan_2020_NonlocalHDR}, HDR-GAN \cite{Niu_2021_HDRGAN}, HDR-Transformer \cite{Liu_2022_ContextHDR}, SCTNet \cite{Tel_2023_HDR} and Ours. The red box represents the reference frame. All 3-input methods are fed with frames marked with a red heart symbol.}
  \label{fig:our_2}
\end{figure}

We validate the capability of the proposed iHDR to handle an arbitrary number of inputs on our collected dataset and compare it with other 3-input frameworks. Fig. \ref{fig:our_2} demonstrate the visual results.

\end{document}